\documentclass[journal=jacsat,manuscript=article,layout=onecolumn]{achemso}

\usepackage{chemformula} 
\usepackage[T1]{fontenc} 

\author{Lizhou Yang}
\affiliation{School of Emerging Technology, Hefei National Research Center for Physical Sciences at the Microscale, and Hefei National Laboratory, University of Science and Technology of China, Hefei, Anhui 230026, China}

\author{Wenneng Zhao}
\affiliation{Wuhan Institute of Technology, Wuhan, Hubei, 430205, China}

\author{Ye Yang}
\affiliation{Department of Physics, Hefei University of Technology, Hefei, Anhui 230601, China}

\author{Weiwei Chen}
\affiliation{Department of Physics, Hefei University of Technology, Hefei, Anhui 230601, China}

\author{YongKeat Chan}
\affiliation{School of Emerging Technology, Hefei National Research Center for Physical Sciences at the Microscale, and Hefei National Laboratory, University of Science and Technology of China, Hefei, Anhui 230026, China}

\author{Chen-Hao Lv}
\affiliation{Wuhan Institute of Technology, Wuhan, Hubei, 430205, China}

\author{Qing-Bo Liu}
\affiliation{Wuhan Institute of Technology, Wuhan, Hubei, 430205, China}
\email{phyqboliu@wit.edu.cn}

\author{Yujia Xiang}
\affiliation{Key Laboratory of Precision and Intelligent Chemistry, University of Science and Technology of China, Hefei, Anhui 230026, China}

\author{Xinming Qin}
\affiliation{School of Emerging Technology, Hefei National Research Center for Physical Sciences at the Microscale, and Hefei National Laboratory, University of Science and Technology of China, Hefei, Anhui 230026, China}

\author{Zedong Li}
\affiliation{Hanhai Lingchao Intelligent Computing (Shanghai) Technology Services Co., Ltd.}

\author{Weiyi Wang}
\affiliation{School of Emerging Technology, Hefei National Research Center for Physical Sciences at the Microscale, and Hefei National Laboratory, University of Science and Technology of China, Hefei, Anhui 230026, China}
\email{dancingw@ustc.edu.cn}

\author{Jinlong Yang}
\affiliation{Key Laboratory of Precision and Intelligent Chemistry, University of Science and Technology of China, Hefei, Anhui 230026, China}

\author{Wei Hu}
\affiliation{Key Laboratory of Precision and Intelligent Chemistry, University of Science and Technology of China, Hefei, Anhui 230026, China}
\email{whuustc@ustc.edu.cn}

\title{High-Throughput Discovery of Chiral Carbon–Boron Semimetals with Weyl Topology and Phonon-Mediated Superconductivity}

\begin{document}

\begin{abstract}
Chiral crystals provide a natural platform for unconventional topological fermions, while light-element carbon--boron frameworks are promising for phonon-mediated superconductivity owing to strong covalent bonding and high phonon frequencies. Here, we combine Wyckoff-position-based structure generation with high-throughput first-principles calculations to search for C--B crystals in selected hexagonal chiral space groups. Starting from 972 generated $\mathrm{C}_m\mathrm{B}_n$ structures ($m,n=1\text{--}4$), successive filters based on energetic stability, dynamical stability, electronic metallicity, topological band features, and thermal stability identify six stable candidates: $(\mathrm{C}_3\mathrm{B}_2)_8$, $(\mathrm{C}\mathrm{B}_2)_3$, $(\mathrm{C}\mathrm{B}_4)_6$, $(\mathrm{C}_3\mathrm{B}_2)_6$, $(\mathrm{C}_2\mathrm{B}_3)_6$, and $(\mathrm{C}\mathrm{B}_2)_6$. These materials host diverse topological semimetal states near the Fermi level, including charge-one, charge-two, and charge-three Weyl points, nodal rings, and nodal-surface states. Notably, $(\mathrm{C}_3\mathrm{B}_2)_8$ exhibits screw-symmetry-protected charge-three Weyl points and strong electron--phonon coupling, yielding a McMillan transition temperature of $31.1\,\mathrm{K}$ for the Coulomb pseudopotential parameter $\mu^* = 0.10$. In addition, $(\mathrm{C}\mathrm{B}_2)_3$ hosts multiple charge-one Weyl points and pronounced anisotropic multiband superconductivity revealed by fully anisotropic Eliashberg calculations. These results extend the materials scope of topological semimetals with phonon-mediated superconductivity to chemically designed chiral light-element structures.

\end{abstract}

\section{1. Introduction}\label{sec:Introduction}

Topological semimetals are gapless quantum materials in which valence and conduction bands cross at symmetry-protected nodes in momentum space.\cite{annurev:/content/journals/10.1146/annurev-matsci-070218-010049,RevModPhys.90.015001} The nontrivial Berry curvature associated with these band crossings gives rise to characteristic phenomena such as Fermi-arc surface states, anomalous transport responses, and nonlinear optical effects.\cite{RevModPhys.93.025002,10.1038/s41578-021-00301-3,PhysRevX.5.031023,10.1038/s41567-018-0234-5,10.1038/s41467-018-06088-2,10.1038/nphys3969,10.1038/s41563-019-0296-5} Within this broad family, chiral-crystal topological semimetals are of particular interest because the absence of spatial inversion, mirror, and rotoinversion symmetries favors unconventional topological band crossings with nontrivial Chern numbers.\cite{Guoqing2018,doi:10.1126/science.aaf5037} These multifold chiral fermions can carry Chern numbers larger than those of conventional Weyl fermions, producing long Fermi-arc surface states\cite{Schroter2019,PhysRevLett.119.206401} and enhanced Berry-curvature-related responses, including quantized circular photogalvanic and spin Hall effects.\cite{10.1038/ncomms15995,PhysRevB.108.L201404}

Combining chiral topological band structures with superconductivity offers a promising route toward topological superconducting phases and Majorana-related excitations.\cite{10.1038/s41578-021-00336-6,10.1093/nsr/nwy155,PhysRevLett.129.027001,PhysRevLett.120.067003} Chiral lattices are especially attractive because they can host unconventional topological crossings, including charge-one, charge-two, and charge-three Weyl points as well as nodal-surface states (Fig. 1(a)). Although high-throughput searches based on topological quantum chemistry and symmetry indicators have greatly advanced the discovery of topological materials,\cite{10.1038/nature23268,10.1038/s41586-019-0954-4,10.1038/s41586-019-0944-6,10.1126/science.abg9094} existing studies mainly focus on known crystalline semimetals and rarely explore chemically designed light-element chiral frameworks. To date, systematic searches for chiral carbon--boron frameworks that integrate diverse Weyl, nodal-ring, and nodal-surface topologies with phonon-mediated superconductivity remain largely unexplored. Strong C--B covalent bonding can support high phonon frequencies and appreciable electron--phonon coupling (Fig. 1(b)).\cite{PhysRevLett.93.237003,PhysRevLett.101.016401,C2NR12018F,PhysRevB.102.024519,PhysRevB.109.184517,PhysRevB.109.144509} Accordingly, chiral C--B compounds warrant a systematic search for the coexistence of symmetry-enforced topological band crossings and appreciable phonon-mediated superconductivity.

Here, we perform a high-throughput first-principles screening of chiral \(\mathrm{C}_m\mathrm{B}_n\) crystals in SGs 168--173 and 177--180. Starting from 972 Wyckoff-generated structures with stoichiometries \(\mathrm{C}_m\mathrm{B}_n\) (\(m,n=1\text{--}4\)) and no more than 40 atoms per unit cell, we apply successive filters based on energetic stability, dynamical stability, metallicity, clean low-energy band crossings near the Fermi level, and thermal stability to identify six stable chiral C--B candidates: \((\mathrm{C}_3\mathrm{B}_2)_8\), \((\mathrm{C}\mathrm{B}_2)_3\), \((\mathrm{C}\mathrm{B}_4)_6\), \((\mathrm{C}_3\mathrm{B}_2)_6\), \((\mathrm{C}_2\mathrm{B}_3)_6\), and \((\mathrm{C}\mathrm{B}_2)_6\). These semimetals host diverse topological states near the Fermi level, including charge-one, charge-two, and charge-three Weyl points (WPs), nodal rings, and nodal-surface states. Among these candidates, \((\mathrm{C}_3\mathrm{B}_2)_8\) exhibits screw-symmetry-protected charge-three WPs and strong electron--phonon coupling with a McMillan transition temperature of \(31.1\,\mathrm{K}\) for the Coulomb pseudopotential parameter \(\mu^* = 0.10\). Moreover, \((\mathrm{C}\mathrm{B}_2)_3\) displays nontrivial Weyl topology together with pronounced anisotropic multiband superconductivity. By extending the search beyond known crystal structures, this work identifies a family of chiral light-element compounds in which phonon-mediated superconductivity coexists with diverse Weyl and nodal states.

\begin{figure*}[!htb]
\begin{center}
\includegraphics[width=1\textwidth]{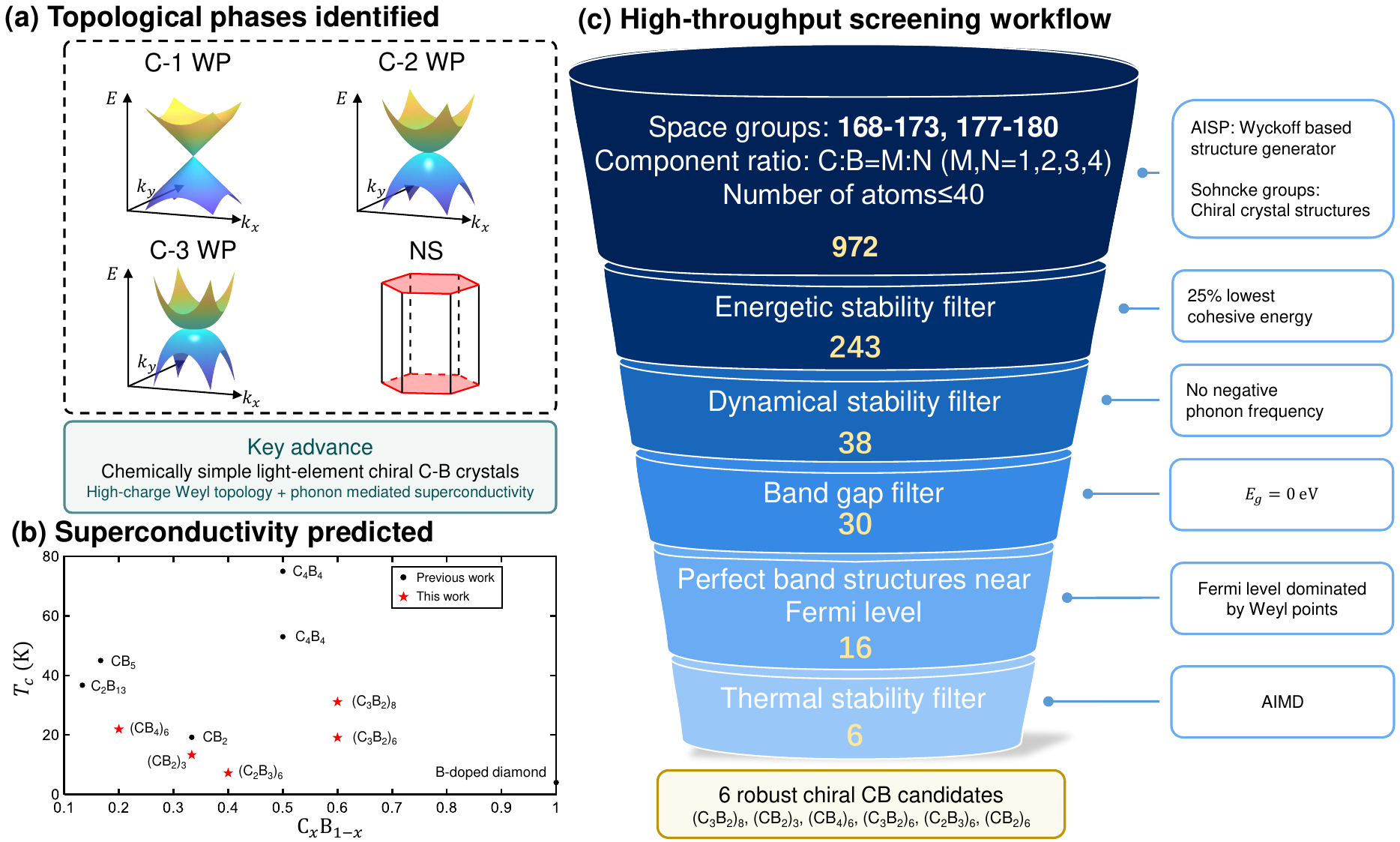}
\end{center}
\caption{
High-throughput screening and topological classification of chiral \(\mathrm{C}_m\mathrm{B}_n\) semimetals. (a) Schematic illustrations of the topological phases identified in the screened materials, including \(|C|=1\), \(|C|=2\), and \(|C|=3\) WPs, as well as nodal-surface (NS) states.
(b) Comparison of the superconducting transition temperatures \(T_{\mathrm c}\) at ambient pressure between previously reported C--B superconductors and the superconducting candidates identified in this work. For consistency, all \(T_{\mathrm c}\) values are estimated using the McMillan formula with \(\mu^\ast = 0.1\)\cite{PhysRevB.69.224505,C2NR12018F,PhysRevLett.93.237003,PhysRevLett.101.016401,PhysRevB.102.024519}.
(c) Screening workflow for nonmagnetic chiral carbon--boron crystals in hexagonal SGs 168--173 and 177--180, with stoichiometries \(m,n=1,2,3,4\) and no more than 40 atoms per unit cell.
Starting from 972 Wyckoff-generated structures, successive filters based on energetic stability, dynamical stability, metallicity (\(E_g=0\)), clean band features near the Fermi level, and thermal stability reduce the candidates to six robust semimetals.}
\label{fig:toc}
\end{figure*}

\section{2. Results and discussion}\label{sec:results}

\subsection{2.1 Representative \(k\cdot p\) models}\label{subsec:kp_models}

Before discussing the six semimetals, we first summarize the minimal
symmetry-constrained \(k\cdot p\) models that describe the representative
topological band crossings identified in the screening. In all cases, the
Hamiltonian is written in a two-band basis as
\begin{equation}
    H_{\mathrm{eff}}(\mathbf k)=\epsilon_0(\mathbf k)\sigma_0+
    \mathbf d(\mathbf k)\cdot\boldsymbol\sigma,
\end{equation}
where \(\mathbf k=(k_x,k_y,k_z)\), \(\sigma_0\) is the identity matrix, and \(\epsilon_0(\mathbf k)\) gives an
overall energy shift. Hermiticity requires \(\epsilon_0(\mathbf k)\) and the
components of \(\mathbf d(\mathbf k)=(d_x(\mathbf{k}),d_y(\mathbf{k}),d_z(\mathbf{k}))\) to be real. The vector
\(\mathbf d(\mathbf k)\) determines the band splitting and eigenstates, with
\(\boldsymbol\sigma=(\sigma_x,\sigma_y,\sigma_z)\) denoting Pauli matrices in
the two-band subspace. The complete fitted \(k\cdot p\) Hamiltonians and
numerical parameters are provided in the Supporting Information; here we retain
only the lowest-order terms required to determine the topology.

\paragraph{2.1.1 Charge-one WP.}
A representative charge-one WP is found on the \(\Gamma-M\) line of the
chiral SG \(P6_2 22\) (No.~180). We denote the crossing point by
\(\mathbf k_0=(u_0,0,0)\), with \(-0.5<u_0<0.5\), where the two crossing bands are assigned to the \(\Sigma_1\) and
\(\Sigma_2\) irreducible representations (IRs) of the little group, respectively.
For the twofold screw operation \(g=\{2_{210}|0,0,1/3\}\), the corresponding
characters are \(+1\) and \(-1\). Thus, in the two-state subspace ordered as
\((\Sigma_1,\Sigma_2)\), one may choose
\begin{equation}
    D(g)=
    \begin{pmatrix}
    1 & 0\\
    0 & -1
    \end{pmatrix}
    =\sigma_z .
\end{equation}
The opposite screw eigenvalues forbid hybridization on the \(\Gamma-M\) line
and protect the crossing.

Around \(\mathbf k_0\), we choose \(k_x\) along the \(\Gamma-M\) direction.
Under \(g\), \(k_x\rightarrow k_x\), whereas
\((k_y,k_z)\rightarrow(-k_y,-k_z)\). The leading symmetry-allowed Hamiltonian is
therefore
\begin{equation}
    H_{C=1}(\mathbf k)=
    \epsilon_0(\mathbf k)\sigma_0+
    v_x k_x\sigma_z+
    v_y k_y\sigma_x+
    v_z k_z\sigma_y ,
\end{equation}
where \(v_x\), \(v_y\), and \(v_z\) are real constants. This Hamiltonian
describes a charge-one WP with
$C=\mathrm{sgn}(v_x v_y v_z)$, where $\mathrm{sgn}$ denotes the sign function.

\paragraph{2.1.2 Charge-two WP.}
A representative charge-two WP occurs at the \(\Gamma\) point of the chiral SG
\(P6_5\) (No.~170). The twofold degeneracy at \(\Gamma\) is assigned to the
conjugate IRs \(\Gamma_3\oplus\Gamma_5\) of the point
group \(C_6\). For the sixfold screw operation \(g=\{6^+_{001}|0,0,1/6\}\), the
corresponding characters are \(e^{i2\pi/3}\) and \(e^{-i2\pi/3}\). Thus, in the
two-state subspace ordered as \((\Gamma_3,\Gamma_5)\), one may choose
\begin{equation}
    D(g)=
    \begin{pmatrix}
    e^{i2\pi/3} & 0\\
    0 & e^{-i2\pi/3}
    \end{pmatrix}.
\end{equation}
The mismatch of the screw eigenvalues forbids a linear transverse coupling
between the two states, so the leading allowed in-plane coupling is quadratic in
\(k_x\) and \(k_y\).

Around \(\Gamma\), choosing \(k_z\) along the screw axis, the minimal
Hamiltonian can be written as
\begin{equation}
    H_{C=2}(\mathbf k)=
    \epsilon_0(\mathbf k)\sigma_0+
    v_z k_z\sigma_z+
    \alpha k_+^2\sigma_+
    +\alpha^\ast k_-^2\sigma_- ,
\end{equation}
where \(v_z\) is real, \(\alpha\) is generally complex,
\(k_\pm=k_x\pm ik_y\), and
\(\sigma_\pm=(\sigma_x\pm i\sigma_y)/2\). The quadratic transverse coupling
winds twice around the degeneracy, giving a double-Weyl point with \(|C|=2\).

\paragraph{2.1.3 Charge-three WP.}
A charge-three WP occurs on the \(\Gamma-A\) line of the
chiral SG \(P6_3\) (No.~173). We denote the crossing point by
\(\mathbf k_0=(0,0,w_0)\), with \(-0.5<w_0<0.5\), where the two crossing bands are assigned to the
\(\Delta_3\oplus\Delta_4\) IRs of the little group \(C_6\), and the relevant little-group constraints are generated by
\(g_2=\{2_{001}|0,0,1/2\}\) and
\(g_3=\{3^+_{001}|0,0,0\}\). In the two-state subspace ordered as
\((\Delta_3,\Delta_4)\), one may choose
\begin{equation}
\begin{aligned}
    D(g_2)&=
    e^{i\pi w_0}
    \begin{pmatrix}
    1 & 0\\
    0 & -1
    \end{pmatrix},\\
    D(g_3)&=
    e^{i2\pi/3}
    \begin{pmatrix}
    1 & 0\\
    0 & 1
    \end{pmatrix}.
\end{aligned}
\end{equation}

Symmetry restrictions along the $\Gamma-A$ axis dictate the off-diagonal coupling: $g_2$ inverts the transverse momentum $(k_x,k_y)\to(-k_x,-k_y)$, requiring an odd functional dependence, and $g_3$ (a $120^\circ$ rotation in $k$-space, $k_\pm\to e^{\pm i2\pi/3}k_\pm$) forbids both linear and quadratic transverse contributions. Consequently, the lowest-order symmetry-allowed Hamiltonian takes the form
\begin{equation}
    H_{C=3}(\mathbf k)=
    \epsilon_0(\mathbf k)\sigma_0+
    v_z k_z\sigma_z+
    \alpha k_+^3\sigma_+
    +\alpha^\ast k_-^3\sigma_- ,
\end{equation}
where \(v_z\) is a real constant and \(\alpha\) is generally a complex constant. The corresponding
dispersion is
\begin{equation}
    E_{\pm}(\mathbf k)=
    \epsilon_0(\mathbf k)\pm
    \sqrt{v_z^2k_z^2+|\alpha|^2(k_x^2+k_y^2)^3}.
\end{equation}
Thus, the bands disperse linearly along the \(\Gamma-A\) direction but cubically
in the transverse plane. The cubic transverse coupling winds three times around
the degeneracy, giving a charge-three WP.

\subsection{2.2 High-throughput screening and candidate selection}\label{subsec:screening}

We generate chiral C--B crystal structures in the hexagonal Sohncke SGs 168--173 and 177--180 using the in-house AISP package.~\cite{aisp} As illustrated in Fig.~\ref{fig:toc}(c), symmetry-allowed Wyckoff-position combinations are enumerated for small-integer C:B stoichiometries ($m,n=1\text{--}4$) and primitive cells containing no more than 40 atoms. This search space prioritizes chemically simple phases of moderate structural complexity for targeted synthesis and crystallographic characterization. The enumeration yields 972 initial structures.

The screening workflow proceeds through successive filters based on energetic stability, dynamical stability, electronic metallicity, topological band features, and thermal stability. First, all 972 generated structures are fully relaxed, and only the lowest 25\% in cohesive energy are retained, reducing the number of candidates to 243. Second, phonon calculations are performed to examine dynamical stability, and structures with imaginary phonon frequencies are removed, leaving 38 candidates. Third, a band-gap filter is applied to select metallic systems with $E_g=0$, yielding 30 candidates. Fourth, the electronic band structures are further inspected to identify clean low-energy band features near the Fermi level, especially those dominated by Weyl-type crossings, reducing the candidate pool to 16. Finally, \textit{ab initio} molecular dynamics simulations are used as a thermal-stability filter, from which six robust chiral C--B candidates are obtained: $(\mathrm{C}_3\mathrm{B}_2)_8$, $(\mathrm{C}\mathrm{B}_2)_3$, $(\mathrm{C}\mathrm{B}_4)_6$, $(\mathrm{C}_3\mathrm{B}_2)_6$, $(\mathrm{C}_2\mathrm{B}_3)_6$, and $(\mathrm{C}\mathrm{B}_2)_6$.

\subsection{2.3 Charge-three Weyl fermions and phonon-mediated superconductivity in \((\mathrm{C}_3\mathrm{B}_2)_8\)}\label{subsec:C24B16}
\begin{figure*}[!htb]
\begin{center}
\includegraphics[width=1\textwidth]{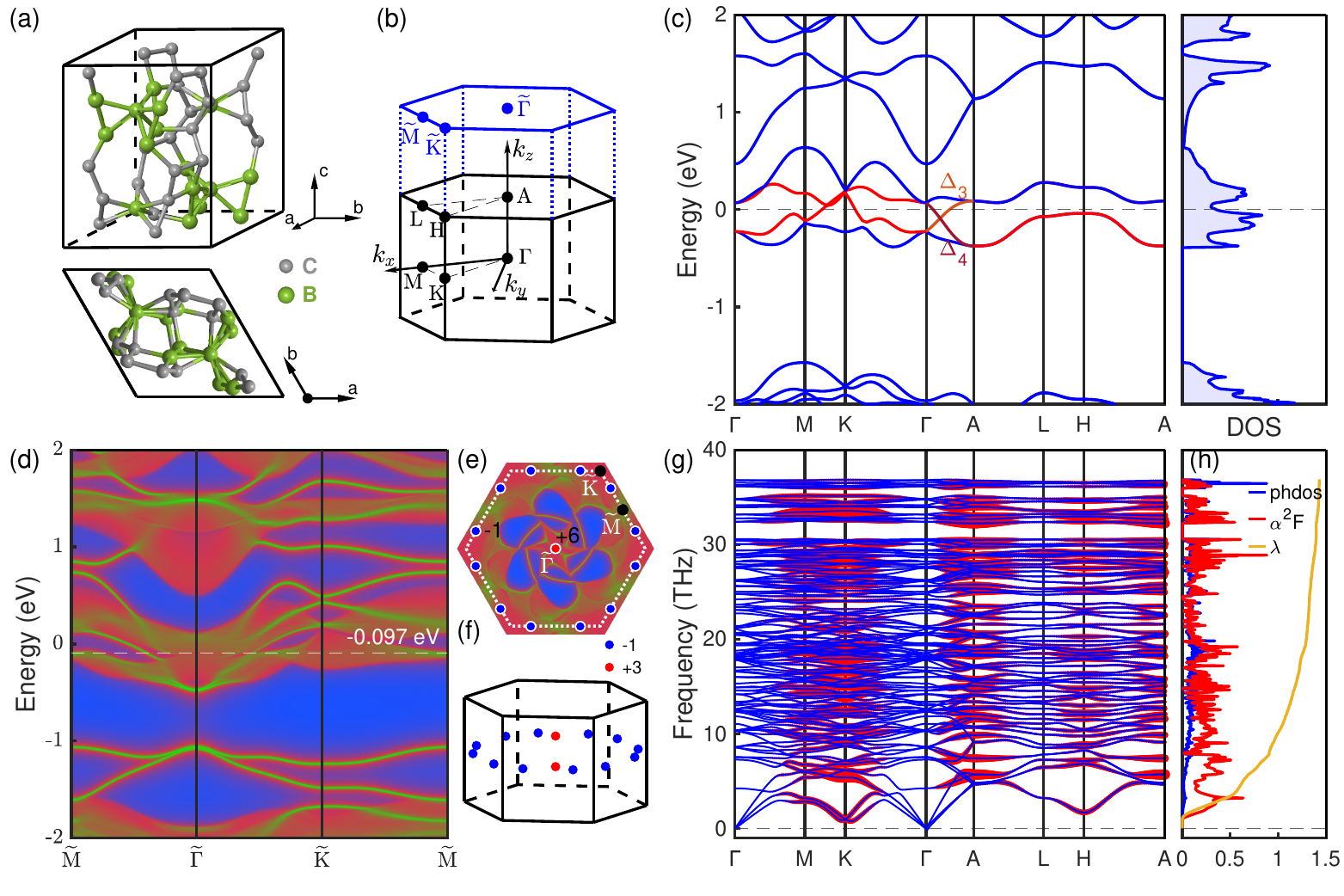}
\end{center}
\caption{
Crystalline, electronic, phononic, superconducting, and topological properties
of \((\mathrm{C}_3\mathrm{B}_2)_8\).
(a) Crystal structure viewed from two representative directions. Gray and green
spheres denote C and B atoms, respectively.
(b) Bulk Brillouin zone and high-symmetry path used for the band-structure
calculations; the blue hexagon denotes the projected \((001)\) surface
Brillouin zone.
(c) Electronic band structure and density of states. The red bands highlight the
low-energy bands forming the topological crossing near the Fermi level, with
\(\Delta_3\) and \(\Delta_4\) labeling the corresponding IRs on the
\(\Gamma-A\) line.
(d) \((001)\) surface spectral function along the projected high-symmetry path.
The dashed line marks the energy \(-0.097\) eV used for the constant-energy
contour in panel (e).
(e) Constant-energy surface contour at \(-0.097\) eV, showing Fermi-arc surface
states. The central projection near \(\tilde{\Gamma}\) carries total chirality
\(+6\), while the boundary projections originate from \(C=-1\) WPs.
(f) Distribution of WPs in the first Brillouin zone. Blue and red dots denote
WPs with \(C=-1\) and \(C=+3\), respectively. (g) Phonon dispersion. The width of the red overlay is proportional
to the calculated phonon linewidth.
(h) Normalized projected phonon density of states, Eliashberg spectral function
\(\alpha^2F(\omega)\), and cumulative electron--phonon coupling strength
\(\lambda(\omega)\). The PhDOS is rescaled for comparison with
\(\alpha^2F(\omega)\) and \(\lambda(\omega)\). Direct integration of
\(\alpha^2F(\omega)\) gives \(\lambda=1.43\), yielding
\(T_c^{\mathrm{McM}}=31.1\,\mathrm{K}\) for \(\mu^\ast = 0.10\).
}
\label{fig:C24B16}
\end{figure*}

Among the six screened candidates, $(\mathrm{C}_3\mathrm{B}_2)_8$ is particularly noteworthy because it combines high-Chern-number Weyl topology with strong phonon-mediated superconductivity within a chemically simple light-element framework. As shown in Fig.~\ref{fig:C24B16}(a), the structure crystallizes in the chiral hexagonal space group $P6_3$ (No.~173); the optimized lattice parameters and Wyckoff positions are listed in Table~S1 of the Supporting Information. The electronic band structure reveals symmetry-protected crossings between the $\Delta_3$ and $\Delta_4$ bands along the $\Gamma-A$ direction at $\mathbf{k}=(0,0,\pm0.147)$ [Fig.~\ref{fig:C24B16}(c)]. According to the symmetry analysis and the effective $k\cdot p$ model discussed above, these crossings correspond to charge-three Weyl points. Berry-curvature integration over small enclosing spheres confirms a topological charge of $C=+3$ for each crossing, establishing $(\mathrm{C}_3\mathrm{B}_2)_8$ as a high-Chern-number Weyl semimetal. Because spin--orbit coupling is neglected, all Weyl charges quoted here are per spin channel, and every crossing has a coincident spin copy. The equality of the charges at $\pm\mathbf{K}_0$ follows from time-reversal symmetry: under $\mathcal{T}$, the Berry curvature and the oriented surface element both change sign under $\mathbf{k}\rightarrow-\mathbf{k}$, leaving the enclosed Berry flux invariant (Sec.~S2B of the Supporting Information).~\cite{PhysRevB.90.045130} In addition, charge-one Weyl points carrying $C=-1$ emerge on the $M-K$ boundary, ensuring the overall chirality compensation required by the Nielsen--Ninomiya theorem. The nontrivial bulk topology is further manifested in the (001) surface electronic structure. As shown in Figs.~\ref{fig:C24B16}(d) and \ref{fig:C24B16}(e), the two charge-three Weyl points along the $\Gamma-A$ direction project onto the same surface momentum near $\tilde{\Gamma}$, generating a total projected chirality of $+6$ per spin channel. In contrast, the six charge-one Weyl points project near the surface Brillouin-zone boundary. Consequently, long Fermi-arc surface states emerge and connect projections with opposite chirality, providing a direct manifestation of the underlying high-Chern-number Weyl topology. The locations and chiralities of all Weyl points are summarized in Fig.~\ref{fig:C24B16}(f).

The phonon spectrum of $(\mathrm{C}_3\mathrm{B}_2)_8$ contains no imaginary frequencies throughout the Brillouin zone, indicating its dynamical stability [Fig.~\ref{fig:C24B16}(g)]. A pronounced softening of the acoustic branch appears along the $H-K$ direction and is accompanied by enhanced phonon linewidths, indicating strong electron--phonon interactions.~\cite{PhysRevB.6.2577} To understand the origin of this anomaly, we calculate the static Lindhard susceptibility, which exhibits pronounced enhancements around the $K$ and $H$ points as well as along the $\Gamma-A$ direction (see Sec.~S3C of the Supporting Information). These features suggest that low-energy electronic scattering contributes to a Kohn-anomaly-like softening of the acoustic branch.~\cite{PhysRevLett.2.393} Eigenvector analysis further reveals that the softened mode is dominated by out-of-plane vibrations of the B atoms occupying the Wyckoff (2b) sites, indicating strong and highly mode-selective electron--phonon coupling. The coincidence between enhanced electronic susceptibility and softened phonon modes therefore supports a Kohn-anomaly-like interpretation, while the strong mode dependence of the phonon linewidths indicates that electron--phonon matrix elements also play an essential role.~\cite{PhysRevB.77.165135,PhysRevResearch.5.013218}

The resulting electron--phonon interaction is reflected in the Eliashberg spectral function $\alpha^2F(\omega)$, phonon density of states, and cumulative coupling strength $\lambda(\omega)$ shown in Figs.~\ref{fig:C24B16}(g) and (h). Direct integration of $\alpha^2F(\omega)$ yields a large electron--phonon coupling constant of $\lambda=1.43$, placing $(\mathrm{C}_3\mathrm{B}_2)_8$ in the strong-coupling regime. Using the McMillan equation with $\mu^\ast = 0.10$, we obtain a superconducting transition temperature of $T_c^{\mathrm{McM}}=31.1\,\mathrm{K}$. The sizable \(T_c\) reflects the combined effects of strong electron--phonon coupling and the characteristic phonon-energy scale. As summarized in Fig.~\ref{fig:toc}(b), this value is among the highest predicted ambient-pressure transition temperatures reported for carbon--boron superconductors.~\cite{PhysRevB.69.224505,C2NR12018F,PhysRevLett.93.237003,PhysRevLett.101.016401,PhysRevB.102.024519} Therefore, $(\mathrm{C}_3\mathrm{B}_2)_8$ provides a rare example in which charge-three Weyl fermions, extended Fermi-arc surface states, and phonon-mediated superconductivity coexist within a single material platform.

\begin{figure*}[!htb]
\begin{center}
\includegraphics[width=1\textwidth]{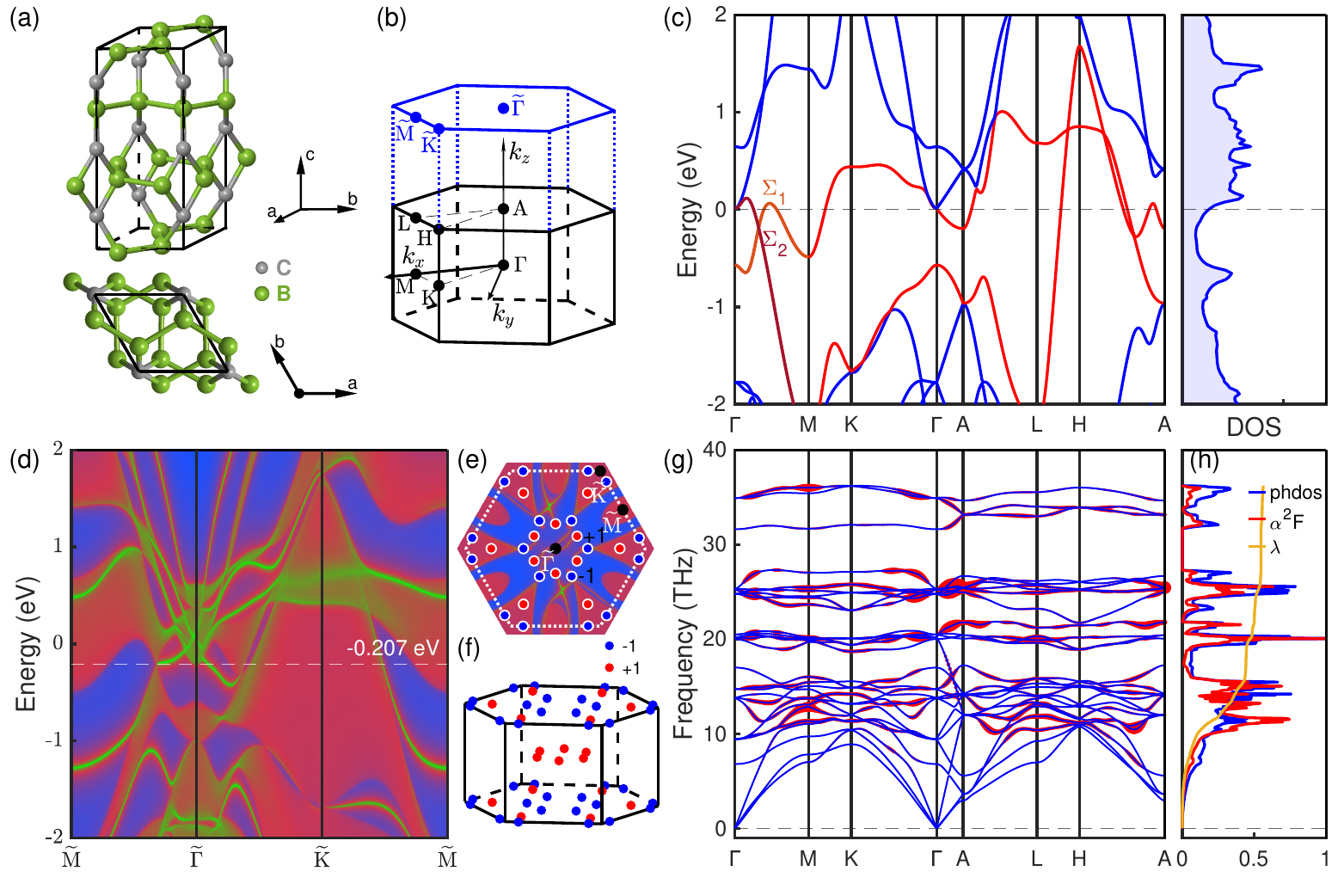}
\end{center}
\caption{
Crystalline, electronic, phononic, superconducting, and topological properties
of \((\mathrm{C}\mathrm{B}_2)_3\).
(a) Crystal structure viewed from two representative directions. Gray and green
spheres denote C and B atoms, respectively.
(b) Bulk Brillouin zone and high-symmetry path used for the band-structure
calculations; the blue hexagon denotes the projected \((001)\) surface
Brillouin zone.
(c) Electronic band structure and density of states. The red bands highlight the
low-energy bands forming the Weyl crossings near the Fermi level, with
\(\Sigma_1\) and \(\Sigma_2\) labeling the corresponding IRs on the
\(\Gamma-M\) line.
(d) \((001)\) surface spectral function along the projected high-symmetry path.
The dashed line marks the energy \(-0.207\) eV used for the constant-energy
contour in panel (e).
(e) Constant-energy surface contour at \(-0.207\) eV, showing Fermi-arc surface states.
(f) Distribution of WPs in the first Brillouin zone. Blue and red dots denote
WPs with \(C=-1\) and \(C=+1\), respectively.
(g) Phonon dispersion. The width of the red overlay is proportional
to the calculated phonon linewidth.
(h) Normalized projected phonon density of states, Eliashberg spectral function
\(\alpha^2F(\omega)\), and cumulative electron--phonon coupling strength
\(\lambda(\omega)\). Direct integration of \(\alpha^2F(\omega)\) gives
\(\lambda=0.56\), yielding \(T_c^{\mathrm{McM}}=11.5\,\mathrm{K}\) for
\(\mu^\ast = 0.10\).
}
\label{fig:C3B6}
\end{figure*}
\subsection{2.4 Charge-one Weyl topology and anisotropic multiband superconductivity in $(\mathrm{C}\mathrm{B}_2)_3$}

Among the screened compounds, $(\mathrm{C}\mathrm{B}_2)_3$ is distinguished by the coexistence of charge-one Weyl topology and pronounced anisotropic multiband superconductivity. As shown in Fig.~\ref{fig:C3B6}(a), it crystallizes in the chiral hexagonal space group $P6_2 22$ (No.~180), and the optimized structural parameters are summarized in Table~S1 of the Supporting Information.
Its electronic structure exhibits symmetry-protected band crossings near the Fermi level [Fig.~\ref{fig:C3B6}(c)]. In particular, the crossing along the $\Gamma-M$ direction involves the $\Sigma_1$ and $\Sigma_2$ irreducible representations of the little group. Because these two bands possess opposite eigenvalues under the screw operation $\{2_{210}|0,0,1/3\}$, they form the $\Sigma_1\oplus\Sigma_2$ subspace discussed above and give rise to symmetry-protected charge-one Weyl points described by the corresponding linear $k\cdot p$ Hamiltonian. The nontrivial topology is further reflected in the (001) surface electronic structure. As shown in Figs.~\ref{fig:C3B6}(d) and \ref{fig:C3B6}(e), arc-like surface states connect projections of Weyl points with opposite chiralities. Compared with $(\mathrm{C}_3\mathrm{B}_2)_8$, the surface-state pattern is more intricate because multiple charge-one Weyl points project onto nearby momenta and partially overlap with bulk-projected states. The corresponding distribution of Weyl points and their chiralities is summarized in Fig.~\ref{fig:C3B6}(f), confirming the Weyl-semimetal nature of $(\mathrm{C}\mathrm{B}_2)_3$.

\begin{figure*}[!htb]
\begin{center}
\includegraphics[width=1\textwidth]{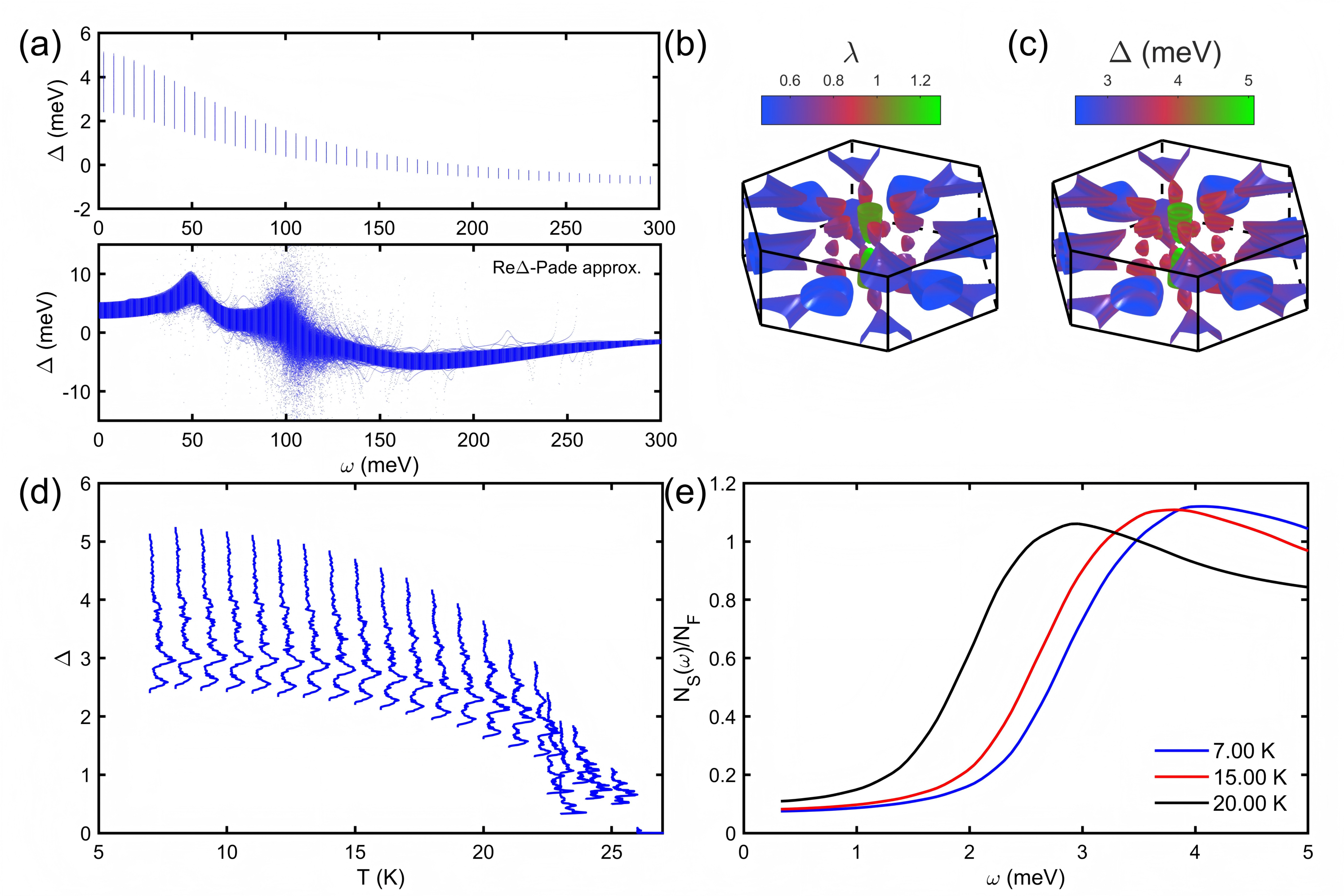}
\end{center}
\caption{
Fully anisotropic superconducting properties of \((\mathrm{C}\mathrm{B}_2)_3\)
obtained from EPW.
(a) Upper panel: superconducting gap on the imaginary-frequency axis at
\(T=10.0\,\mathrm{K}\), obtained by solving the fully anisotropic Eliashberg equations
with \(\mu^\ast = 0.10\). Lower panel: superconducting gap on the real-frequency
axis, obtained from analytic continuation using Pad\'e approximants.
(b) Distribution of the Fermi-surface-resolved electron--phonon coupling
strength \(\lambda_{n\mathbf{k}}\).
(c) Distribution of the anisotropic superconducting gap
\(\Delta_{n\mathbf{k}}\) on the Fermi surface at \(T=10.0\,\mathrm{K}\).
(d) Temperature evolution of the anisotropic superconducting gaps on the Fermi
surface. The gap distribution is suppressed near \(26\,\mathrm{K}\).
(e) Corresponding quasiparticle density of states
\(N_{\mathrm s}(\omega)/N_{\mathrm F}\) at representative temperatures.
}
\label{fig:C3B6_EPW}
\end{figure*}

The phonon spectrum contains no imaginary frequencies throughout the Brillouin zone, confirming the dynamical stability of the structure [Fig.~\ref{fig:C3B6}(g)]. The calculated phonon linewidths indicate momentum- and mode-dependent electron--phonon interactions. The Eliashberg spectral function $\alpha^2F(\omega)$, phonon density of states, and cumulative coupling strength $\lambda(\omega)$ are presented in Fig.~\ref{fig:C3B6}(h). Integration of $\alpha^2F(\omega)$ yields an electron--phonon coupling constant of $\lambda=0.56$, corresponding to a McMillan transition temperature of $T_c^{\mathrm{McM}}=11.5\,\mathrm{K}$ for $\mu^\ast = 0.10$. To resolve the momentum and band dependence of the superconducting state, we solve the fully anisotropic Migdal--Eliashberg equations using \textsc{EPW}. The momentum-resolved electron--phonon coupling strength $\lambda_{n\mathbf{k}}$ exhibits pronounced variations across the multiband Fermi surface [Fig.~\ref{fig:C3B6_EPW}(b)], leading to a strongly anisotropic superconducting gap distribution $\Delta_{n\mathbf{k}}$ [Fig.~\ref{fig:C3B6_EPW}(c)]. The superconducting gap gradually decreases with increasing temperature and is nearly suppressed around $26\,\mathrm{K}$ [Fig.~\ref{fig:C3B6_EPW}(d)], as further reflected by the evolution of the quasiparticle density of states [Fig.~\ref{fig:C3B6_EPW}(e)]. This transition temperature is substantially higher than the isotropic McMillan estimate, demonstrating that anisotropic and multiband pairing effects play a crucial role in the superconductivity of $(\mathrm{C}\mathrm{B}_2)_3$.

To examine the effect of screened electron--electron interactions, we perform an independent superconducting density functional theory calculation using \textsc{SCTK}. On the adopted $k$- and $q$-point meshes, SCTK yields $T_c^{\mathrm{SCTK}}=8.1\,\mathrm{K}$, providing complementary evidence for phonon-mediated superconductivity when the screened Coulomb interaction is treated explicitly. The SCTK results further indicate that the screened Coulomb repulsion is strongly momentum dependent and can partially suppress the superconducting gap in regions with strong electron--phonon coupling. Overall, $(\mathrm{C}\mathrm{B}_2)_3$ represents a rare chiral material in which charge-one Weyl fermions coexist with anisotropic multiband phonon-mediated superconductivity.

Within the weak-pairing framework, the topology of a fully gapped
time-reversal-invariant superconducting state is controlled jointly by the
Chern numbers of the normal-state Fermi surfaces and the relative signs of
their gaps. For such a state,
\begin{equation}
N_W=\frac{1}{2}\sum_s \operatorname{sgn}(\delta_s)C_{1s},
\end{equation}
where \(C_{1s}\) is the Chern number of Fermi surface \(s\).~\cite{PhysRevB.81.134508}
Here, \(\delta_s\) denotes the time-reversal-projected superconducting gap on
Fermi surface \(s\), i.e., the intraband amplitude that pairs a Bloch state
with its time-reversed partner; its sign determines the relative \(0\) and
\(\pi\) phases between different Fermi surfaces.
For the present system, both EPW and SCDFT yield the same sign of the gap on
the relevant Fermi surfaces, leading to \(N_W=0\) (see Sec.~S3G).
Although such phase-reversed pairing is not obtained within the present
electron--phonon coupling framework, it may be realized through additional
pairing interactions, including repulsive inter-Fermi-surface scattering,
competition between interunit-cell attraction and on-site Coulomb repulsion,
or vertex-corrected electron--phonon coupling in nested, nonadiabatic regimes.
\cite{PhysRevB.90.045130,PhysRevB.103.L161113,PhysRevB.104.L140506}

The other four candidates realize distinct combinations of Weyl and nodal topology: $(\mathrm{C}\mathrm{B}_4)_6$ exhibits a charge-two Weyl phase accompanied by nodal-surface states, $(\mathrm{C}_3\mathrm{B}_2)_6$ hosts charge-three Weyl points and nodal rings, $(\mathrm{C}_2\mathrm{B}_3)_6$ exhibits symmetry-related charge-one Weyl points, and $(\mathrm{C}\mathrm{B}_2)_6$ combines charge-one Weyl points with nodal-surface states (Figs.~S4--S7). Collectively, the six compounds show that charge-one, charge-two, and charge-three Weyl points can occur together with nodal rings and nodal surfaces within a single family of chiral light-element compounds.

\section*{3. Conclusions and Outlook}

In summary, our high-throughput first-principles search identifies six dynamically
and thermally stable chiral carbon--boron compounds from 972 Wyckoff-generated
structures. These light-element frameworks host diverse topological semimetal
states, including charge-one, charge-two, and charge-three Weyl points, nodal
rings, nodal-surface states, and the associated Fermi-arc surface states.
Among them, $(\mathrm{C}_3\mathrm{B}_2)_8$ exhibits screw-symmetry-protected
charge-three Weyl points together with strong electron--phonon coupling, yielding
a calculated superconducting transition temperature of
$T_c^{\mathrm{McM}}=31.1\,\mathrm{K}$. In addition, fully anisotropic Eliashberg
calculations for $(\mathrm{C}\mathrm{B}_2)_3$ reveal pronounced multiband
superconductivity with strongly momentum-dependent gap structures.

Our results demonstrate that chiral carbon--boron compounds provide a versatile
platform for exploring the interplay between crystalline chirality, Weyl topology,
and phonon-mediated superconductivity. While the conventional electron--phonon
pairing states obtained from EPW and SCDFT are topologically trivial due to the
absence of relative gap-sign reversal between Fermi surfaces, nontrivial
topological superconductivity may emerge from additional pairing interactions
beyond the conventional EPC framework. These findings motivate further studies
of unconventional gap structures, interaction effects, and surface
superconductivity in chemically designed chiral light-element materials.

\section{4. Computational method}

First-principles calculations are performed using VASP~\cite{vasp} with the projector augmented-wave method~\cite{blochl1994projector} and the PBE exchange--correlation functional.~\cite{perdew1996generalized} A plane-wave cutoff of 400 eV and Monkhorst--Pack $k$-point meshes with a reciprocal-space resolution of $2\pi\times0.04~\text{\AA}^{-1}$ are used for the high-throughput screening, while the final candidates are recalculated with a cutoff of 500 eV. Structures are first relaxed until the residual forces are below $0.04~\mathrm{eV}/\text{\AA}$ and the total-energy change is below $10^{-5}~\mathrm{eV}$, followed by stricter relaxations with thresholds of $0.02~\mathrm{eV}/\text{\AA}$ and $10^{-8}~\mathrm{eV}$. Space groups are confirmed using \textsc{spglib},~\cite{spglib} and duplicate structures are removed using \textsc{pymatgen}.~\cite{pymatgen} Dynamical stability is examined by DFPT phonon calculations using \textsc{Phonopy}. ~\cite{phonopy-phono3py-JPCM} Thermal and mechanical stabilities are assessed using 5 ps \textit{ab initio} molecular dynamics simulations and elastic-constant calculations, respectively.~\cite{PhysRevB.90.224104}

Spin--orbit coupling is neglected throughout. The electronic bands are therefore twofold spin degenerate, and all Weyl charges, Berry-flux integrals, and projected surface chiralities are reported per spin channel.

Band irreducible representations are obtained using \textsc{Irvsp}.~\cite{GAO2021107760} Maximally localized Wannier functions are constructed with \textsc{Wannier90},~\cite{Pizzi2020} and the resulting tight-binding models are used to calculate Chern numbers and surface spectral functions using \textsc{WannierTools}.~\cite{WU2018405} Effective $k\cdot p$ Hamiltonians are constructed using \textsc{VASP2KP}.~\cite{VASP2KP} Weyl-point chiralities are evaluated by Berry-curvature integration over small enclosing spheres following the \textsc{Z2Pack} scheme~\cite{PhysRevB.95.075146} and cross-checked with Wannier tight-binding calculations.

Electron--phonon coupling calculations are performed using DFPT as implemented in \textsc{Quantum ESPRESSO}.~\cite{Giannozzi_2009,Giannozzi_2017} Optimized norm-conserving Vanderbilt pseudopotentials~\cite{PhysRevB.88.085117} and a kinetic-energy cutoff of 90 Ry are adopted. Self-consistent calculations use $k$-point meshes with a reciprocal-space resolution of $2\pi\times0.03~\text{\AA}^{-1}$, and phonon calculations use $q$-point meshes with a resolution of $2\pi\times0.06~\text{\AA}^{-1}$. The electron--phonon coupling constant $\lambda$, logarithmic average phonon frequency $\omega_{\ln}$, and McMillan transition temperature $T_c^{\mathrm{McM}}$ are obtained from $\alpha^2F(\omega)$ with $\mu^\ast = 0.10$. For $(\mathrm{C}\mathrm{B}_2)_3$, fully anisotropic Migdal--Eliashberg calculations are carried out using \textsc{EPW}~\cite{epw} on interpolated $72\times72\times24$ electronic and $36\times36\times12$ phonon grids. The real-axis gap is obtained by Pad\'e analytic continuation. The effects of the screened Coulomb interaction are further examined using \textsc{SCTK}~\cite{PhysRevB.101.134511} on $12\times12\times4$ electronic and $6\times6\times2$ phonon meshes.

\begin{acknowledgement}

This work is partly supported by the National Natural Science Foundation of China (12404085, 22303098, 42550106, 22322304, and 12595330), Shanghai City's 2024 Targeted Commissioned Project in the Field of Quantum Technology, the Innovation Program for Quantum Science and Technology (2021ZD0303306), the Strategic Priority Research Program of the Chinese Academy of Sciences (XDB1170000 and XDB0450101), the Anhui Province Science and Technology Innovation Project (202423k09020010), the Anhui Initiative in Quantum Information Technologies (AHY090400), the Laoshan Laboratory (LSKJ202300305), the Fundamental Research Funds for the Central Universities (No. WK9990000152), USTC Tang Scholar, the University of Science and Technology of China--Southwest University of Science and Technology Counterpart Cooperation and Development Joint Fund (KY2490002501), and the Shiyan Key Laboratory of Quantum Information and Precision Optics (Hubei University of Automotive Technology) (No. SYZDK12024B03). The AI-driven experiments, simulations, and model training were performed on the robotic AI-Scientist platform of the Chinese Academy of Sciences. The authors thank the Hefei Advanced Computing Center, the Supercomputing Center of the Chinese Academy of Sciences, the Supercomputing Center of USTC, the National Supercomputing Center in Wuxi, and the Tianjin, Shanghai, and Guangzhou Supercomputing Centers for computational resources and support.
\end{acknowledgement}



\begin{suppinfo}
Supplementary information.
\end{suppinfo}

\bibliography{references}

\end{document}